\documentclass[twocolumn,iop]{emulateapj}

\usepackage{footnote}
\usepackage[breaklinks,colorlinks,urlcolor=blue,citecolor=blue,linkcolor=blue]{hyperref}

\pdfoutput=1

\def\Brown{\altaffilmark{1}}
\def\Browntxt{\altaffiltext{1}{Department of Physics, Brown University, Providence, RI 02912, USA}}

\def\UW{\altaffilmark{2}}
\def\UWtxt{\altaffiltext{2}{Department of Physics, University of Washington, Seattle, WA 98195, USA}}

\def\ASU{\altaffilmark{3}}
\def\ASUtxt{\altaffiltext{3}{School of Earth and Space Exploration, Arizona State University, Tempe, AZ 85287, USA}}

\def\USydney{\altaffilmark{4}}
\def\USydneytxt{\altaffiltext{4}{Sydney Institute for Astronomy, School of Physics, The University of Sydney, NSW 2006, Australia}}

\def\SKASA{\altaffilmark{5}}
\def\SKASAtxt{\altaffiltext{5}{Square Kilometre Array South Africa (SKA SA), Pinelands 7405, South Africa}}

\def\Rhodes{\altaffilmark{6}}
\def\Rhodestxt{\altaffiltext{6}{Department of Physics and Electronics, Rhodes University, Grahamstown 6140, South Africa}}

\def\CfA{\altaffilmark{7}}
\def\CfAtxt{\altaffiltext{7}{Harvard-Smithsonian Center for Astrophysics, Cambridge, MA 02138, USA}}

\def\Curtin{\altaffilmark{8}}
\def\Curtintxt{\altaffiltext{8}{International Centre for Radio Astronomy Research, Curtin University, Bentley, WA 6102, Australia}}

\def\CAASTRO{\altaffilmark{9}}
\def\CAASTROtxt{\altaffiltext{9}{ARC Centre of Excellence for All-sky Astrophysics (CAASTRO)}}

\def\ANU{\altaffilmark{10}}
\def\ANUtxt{\altaffiltext{10}{Research School of Astronomy and Astrophysics, Australian National University, Canberra, ACT 2611, Australia}}

\def\Haystack{\altaffilmark{11}}
\def\Haystacktxt{\altaffiltext{11}{MIT Haystack Observatory, Westford, MA 01886, USA}}

\def\Kavli{\altaffilmark{12}}
\def\Kavlitxt{\altaffiltext{12}{Kavli Institute for Astrophysics and Space Research, Massachusetts Institute of Technology, Cambridge, MA 02139, USA}}

\def\RRI{\altaffilmark{13}}
\def\RRItxt{\altaffiltext{13}{Raman Research Institute, Bangalore 560080, India}}

\def\UCB{\altaffilmark{14}}
\def\UCBtxt{\altaffiltext{14}{Department of Astronomy, University of California Berkeley, Berkeley, CA 94720, USA}}

\def\BCCP{\altaffilmark{15}}
\def\BCCPtxt{\altaffiltext{15}{Berkeley Center for Cosmological Physics, University of California Berkeley, Berkeley, CA 94720, USA}}

\def\MIT{\altaffilmark{16}}
\def\MITtxt{\altaffiltext{16}{Department of Physics, Massachusetts Institute of Technology, Cambridge, MA 02139, USA}}

\def\Victoria{\altaffilmark{17}}
\def\Victoriatxt{\altaffiltext{17}{School of Chemical \& Physical Sciences, Victoria University of Wellington, Wellington 6140, New Zealand}}

\def\UWisc{\altaffilmark{18}}
\def\UWisctxt{\altaffiltext{18}{Department of Physics, University of Wisconsin--Milwaukee, Milwaukee, WI 53201, USA}}

\def\UMichigan{\altaffilmark{19}}
\def\UMichigantxt{\altaffiltext{19}{Department of Atmospheric, Oceanic and Space Sciences, University of Michigan, Ann Arbor, MI 48109, USA}}

\def\UMelbourne{\altaffilmark{20}}
\def\UMelbournetxt{\altaffiltext{20}{School of Physics, The University of Melbourne, Parkville, VIC 3010, Australia}}

\def\CASS{\altaffilmark{21}}
\def\CASStxt{\altaffiltext{21}{CSIRO Astronomy and Space Science (CASS), PO Box 76, Epping, NSW 1710, Australia}}

\def\Tata{\altaffilmark{22}}
\def\Tatatxt{\altaffiltext{22}{National Centre for Radio Astrophysics, Tata Institute for Fundamental Research, Pune 411007, India}}

\def\NRAO{\altaffilmark{23}}
\def\NRAOtxt{\altaffiltext{23}{National Radio Astronomy Observatory, Charlottesville and Greenbank, USA}}

\def\SKA{\altaffilmark{24}}
\def\SKAtxt{\altaffiltext{24}{SKA Organisation, Jodrell Bank Observatory, Lower Withington, Macclesfield, SK11 9DL, UK}}

\def\NSFFellow{\altaffilmark{25}}
\def\NSFFellowtxt{\altaffiltext{25}{National Science Foundation Astronomy and Astrophysics Postdoctoral Fellow}}

\begin{document}

\title{The Importance of Wide-field Foreground Removal for 21~cm Cosmology: A Demonstration With Early MWA Epoch of Reionization Observations}

\author{
J.~C.~Pober\Brown$^,$\UW$^,$\NSFFellow,
B.~J.~Hazelton\UW,
A.~P.~Beardsley\ASU$^,$\UW,
N.~A.~Barry\UW,
Z.~E.~Martinot\UW,
I.~S.~Sullivan\UW,
M.~F.~Morales\UW,
M.~E.~Bell\USydney,
G.~Bernardi\SKASA$^,$\Rhodes$^,$\CfA,
N.~D.~R.~Bhat\Curtin$^,$\CAASTRO,
J.~D.~Bowman\ASU,
F.~Briggs\ANU,
R.~J.~Cappallo\Haystack,
P.~Carroll\UW, 
B.~E.~Corey\Haystack,
A.~de~Oliveira-Costa,\Kavli
A.~A.~Deshpande\RRI,
Joshua.~S.~Dillon\UCB$^,$\BCCP$^,$\MIT,
D.~Emrich\Curtin,
A.~M.~Ewall-Wice\MIT,
L.~Feng\MIT$^,$\Kavli,
R.~Goeke\Kavli,
L.~J.~Greenhill\CfA,
J.~N.~Hewitt\MIT$^,$\Kavli,
L.~Hindson\Victoria,
N.~Hurley-Walker\Curtin,
D.~C.~Jacobs\ASU$^,$\NSFFellow,
M.~Johnston-Hollitt\Victoria,
D.~L.~Kaplan\UWisc, 
J.~C.~Kasper\CfA$^,$\UMichigan,
Han-Seek~Kim\UMelbourne$^,$\CAASTRO,
P.~Kittiwisit\ASU,
E.~Kratzenberg\Haystack,
N.~Kudryavtseva\Curtin,
E.~Lenc\USydney$^,$\CAASTRO,
J.~Line\UMelbourne$^,$\CAASTRO,
A.~Loeb\CfA,
C.~J.~Lonsdale\Haystack, 
M.~J.~Lynch\Curtin,
B.~McKinley\ANU, 
S.~R.~McWhirter\Haystack,
D.~A.~Mitchell\CAASTRO$^,$\CASS, 
E.~Morgan\Kavli,
A.~R.~Neben\MIT,
D.~Oberoi\Tata,
A.~R.~Offringa\ANU$^,$\CAASTRO, 
S.~M.~Ord\Curtin$^,$\CAASTRO,
Sourabh~Paul\RRI,
B.~Pindor\UMelbourne$^,$\CAASTRO,
T.~Prabu\RRI,
P.~Procopio\UMelbourne,
J.~Riding\UMelbourne,
A.~E.~E.~Rogers\Haystack, 
A.~Roshi\NRAO, 
Shiv~K.~Sethi\RRI, 
N.~Udaya~Shankar\RRI,
K.~S.~Srivani\RRI, 
R.~Subrahmanyan\RRI$^,$\CAASTRO,
M.~Tegmark\MIT,
Nithyanandan~Thyagarajan\ASU, 
S.~J.~Tingay\Curtin$^,$\CAASTRO,
C.~M.~Trott\Curtin$^,$\CAASTRO,
M.~Waterson\SKA,
R.~B.~Wayth\Curtin$^,$\CAASTRO, 
R.~L.~Webster\UMelbourne$^,$\CAASTRO,
A.~R.~Whitney\Haystack, 
A.~Williams\Curtin, 
C.~L.~Williams\MIT,
J.~S.~B.~Wyithe\UMelbourne$^,$\CAASTRO
}

\Browntxt
\UWtxt
\ASUtxt
\USydneytxt
\SKASAtxt
\Rhodestxt
\CfAtxt
\Curtintxt
\CAASTROtxt
\ANUtxt
\Haystacktxt
\Kavlitxt
\RRItxt
\UCBtxt
\BCCPtxt
\MITtxt
\Victoriatxt
\UWisctxt
\UMichigantxt
\UMelbournetxt
\CASStxt
\Tatatxt
\NRAOtxt
\SKAtxt
\NSFFellowtxt

\begin{abstract}

In this paper we present observations, simulations, and analysis 
demonstrating the direct connection between the location of foreground
emission on the sky and its location in cosmological power spectra from interferometric
redshifted 21~cm experiments.  We begin with a heuristic formalism
for understanding the mapping of sky coordinates into the cylindrically
averaged power spectra measurements used by 21~cm experiments, with
a focus on the effects of the instrument beam response and the
associated sidelobes.
We then demonstrate this mapping by analyzing power spectra with both simulated and
observed data from
the Murchison Widefield Array.  We find that removing a foreground model
which includes sources in both the main field-of-view and the first sidelobes
reduces the contamination in high $k_{\parallel}$ modes by several percent
relative to a model which only includes sources in the main field-of-view,
with the completeness of the foreground model setting the principal
limitation on the amount of power removed.
While small, a percent-level amount of foreground power is in itself more than 
enough to prevent recovery of any EoR signal from these modes.
This result demonstrates that foreground subtraction for redshifted 21~cm
experiments is truly a wide-field problem, and algorithms and simulations
must extend beyond the main instrument field-of-view to potentially
recover the full 21~cm power spectrum.

\end{abstract}


\section{Introduction}

A major goal of modern experimental cosmology is the detection of 21~cm 
emission from neutral hydrogen at high redshifts.  Depending on the redshifts
studied, these observations can probe a wide range of physical and
astrophysical phenomena.  Observations at $\sim 100 \mbox{--} 200$~MHz 
($z\sim6-13$ in the 21~cm line) probe the Epoch of Reionization (EoR) ---
the reionization of the intergalactic medium (IGM) by ultraviolet photons
emitted by the first stars and galaxies.  Observations at higher frequencies (lower
redshifts) trace the neutral hydrogen that remains in galactic halos, and
provide a low resolution ``intensity map" of large scale structure and,
potentially,
the Baryon Acoustic Oscillation (BAO) features in the power spectrum.
At lower frequencies (higher redshifts), one begins to trace the birth of
the first stars during ``Cosmic Dawn" and even the preceding Dark Ages.
For reviews of the 21~cm cosmology technique and the associated science
drivers, see \cite{furlanetto_et_al_2006}, \cite{morales_and_wyithe_2011},
\cite{pritchard_and_loeb_2012}, and \cite{zaroubi_2013}.

A large number of experiments seeking
to detect the power spectra of 21~cm fluctuations are already operational
or being commissioned,
including
the LOw Frequency ARray (LOFAR; \citealt{yatawatta_et_al_2013,van_haarlem_et_al_2013})\footnote{http://www.lofar.org}, 
21 CentiMeter Array (21CMA; \citealt{zheng_et_al_2012})\footnote{http://21cma.bao.ac.cn/index.html}, 
the Giant Metrewave Radio Telescope EoR Experiment (GMRT; \citealt{paciga_et_al_2013})\footnote{http://www.ncra.tifr.res.in/ncra/gmrt},
the MIT Epoch of Reionization Experiment (MITEoR; \citealt{zheng_et_al_2014}),
the Donald C. Backer Precision Array for Probing the Epoch of Reionization
(PAPER; \citealt{parsons_et_al_2010})\footnote{http://eor.berkeley.edu},
and the Murchison Widefield Array (MWA; \citealt{lonsdale_et_al_2009,tingay_et_al_2013,bowman_et_al_2013})\footnote{http://www.mwatelescope.org},
all of which are targeting the signal from the EoR.
A number of additional experiments are also under construction or
planned, such as
the low-frequency Square Kilometre Array (SKA-low; \citealt{mellema_et_al_2013})\footnote{http://www.skatelescope.org} 
and the Hydrogen Epoch of Reionization Array (HERA; \citealt{pober_et_al_2014})\footnote{http://reionization.org}
at EoR and Cosmic Dawn redshifts,
and BAOs from Integrated Neutral Gas Observations (BINGO; \citealt{battye_et_al_2013}),
TianLai\footnote{http://tianlai.bao.ac.cn},
BAORadio \citep{ansari_et_al_2012a,ansari_et_al_2012b},
the Canadian Hydrogen Intensity Mapping Experiment (CHIME; \citealt{shaw_et_al_2014})\footnote{http://chime.phas.ubc.ca},
and the BAO Broadband and Broad-beam experiment (BAOBAB; \citealt{pober_et_al_2013a})
at lower redshifts.

At all redshifts, however, 21~cm experiments are limited by both the inherent
faintness of the cosmological signal and the presence of foregrounds
which can exceed the 21~cm emission by as much as 5 orders of magnitude in
brightness temperature \citep{santos_et_al_2005,yatawatta_et_al_2013,bernardi_et_al_2013,pober_et_al_2013b}.  As such,
the only current detection of HI at cosmological distances comes from
cross-correlation studies using maps from the Green Bank Telescope 
and optical galaxy surveys \citep{chang_et_al_2010,masui_et_al_2013,switzer_et_al_2013}.  Analysis techniques for recovering the signal focus on the relative spectral 
smoothness of the foreground emission as an axis for distinguishing these
contaminants from the 21~cm emission.  Over the past decade, a large body
of literature has worked to develop pipelines that can subtract foreground
sources from 21~cm data sets
(e.g. \citealt{morales_et_al_2006}, 
\citealt{bowman_et_al_2009}, \citealt{liu_et_al_2009}, 
\citealt{liu_and_tegmark_2011}, \citealt{chapman_et_al_2012},
\citealt{dillon_et_al_2013}, \citealt{chapman_et_al_2013},
\citealt{wang_et_al_2013}).
More recently, however, studies of the chromatic interaction of an
interferometer with foreground emission have demonstrated that 
smooth spectrum foregrounds
will occupy an anisotropic wedge-like region of cylindrical 
($k_{\perp},k_{\parallel}$) Fourier space, leaving an ``EoR window"
above the wedge where the 21~cm signal can be cleanly observed \citep{datta_et_al_2010,vedantham_et_al_2012,morales_et_al_2012,parsons_et_al_2012b,trott_et_al_2012,thyagarajan_et_al_2013,liu_et_al_2014a,liu_et_al_2014b}.  
These predictions have since been confirmed
in data sets from PAPER and the MWA 
\citep{pober_et_al_2013b,dillon_et_al_2014,parsons_et_al_2014,jacobs_et_al_2015,ali_et_al_2015,thyagarajan_et_al_2015b},
although significantly more
sensitive observations will be necessary to see if the window remains
uncontaminated down to the level of the 21~cm signal.  \cite{pober_et_al_2014}
demonstrate that while current EoR observatories (PAPER, the MWA, and LOFAR)
do not possess the sensitivity to detect the 21~cm signal with this pure
``foreground avoidance" technique, next-generation experiments like HERA and
the SKA-low can yield high fidelity power spectrum measurements using this
approach, and begin to place constraints on the physics of reionization.\footnote{Although \citealt{pober_et_al_2014} 
focused on results from EoR-frequency experiments,
the ``wedge" and ``EoR window" breakdown is generic for all 21~cm studies
\citep{pober_et_al_2013a}.}  
However, the cosmological signal strength peaks on large scales, so that $k$ modes within
the wedge can have significantly more 21~cm power than modes within the window.
\cite{pober_et_al_2014} show
that if foregrounds can be subtracted from 21~cm data sets, allowing the
recovery of $k$ modes from within the wedge, then the significance of any
power spectrum measurement can be substantially boosted --- enabling the current
generation of 21~cm experiments to make a detection.

Continued research into foreground subtraction algorithms is therefore clearly
well motivated.  As of yet, no technique ---
whether subtracting a model of the sky or using a parametrized fit in
frequency ---
has demonstrated that foreground emission in actual observations
can be removed to the thermal noise level of current
instruments
(although the EoR window has to-date proven relatively free of
foregrounds when
care is taken to limit leakage from the wedge
\citep{pober_et_al_2013a,parsons_et_al_2014,jacobs_et_al_2015,ali_et_al_2015}).
The purpose of this work is to investigate some of the
wide-field effects that complicate the removal of foreground emission
using data from the MWA.  In particular, we focus on the contribution of
sources outside the main lobe of the instrument primary beam
(in this work, we use the term primary beam to refer to the
all-sky power pattern of the antenna or tile element, including
sidelobes).  Far from
the pointing center, chromatic effects in the interferometer
response become stronger; sources out in the
sidelobes of the primary beam therefore create
foreground contamination in higher $k_{\parallel}$ modes than
sources near the pointing center.  Here, we explore this effect in more detail.

This paper is structured as follows.
In \S\ref{sec:formalism}, we lay out a heuristic derivation of how the
instrument primary beam enters in measurements of the 21~cm power spectrum
and how foregrounds are distributed throughout the $(k_{\perp},k_{\parallel})$
plane.
In \S\ref{sec:mwa}, we briefly describe the MWA and the data analyzed in this study.
In \S\ref{sec:sims}, we build on the pedagogical nature of the previous analysis
through simulated MWA power spectra using a sky model containing a single point source of emission.
By changing the location of this source, we demonstrate these primary beam effects in a realistic
but controlled fashion.
In \S\ref{sec:data}, we describe
the calibration, pre-processing, and foreground subtraction applied to the observed data
before making a power spectrum.
The main result is presented in \S\ref{sec:pspec}, 
where we compare power spectra made from our data, where we 
have both subtracted a foreground model which includes sources in the beam
sidelobes and one which does not.
We discuss the implications of these results for future foreground subtraction
efforts in \S\ref{sec:discussion} and conclude in \S\ref{sec:conclusions}.

\section{Wide-Field Effects in the EoR Power Spectrum}
\label{sec:formalism}

Although many 21~cm experiments have wide fields of view, only recently have studies focused
on how wide-field effects might complicate measurements of the 21~cm power spectrum.
Theoretical work has identified the foreground wedge described above and provided a formalism
for mapping the position of foreground emission on the sky to $k$ modes of the 21~cm 
power spectrum \citep{vedantham_et_al_2012,morales_et_al_2012,parsons_et_al_2012b,trott_et_al_2012,thyagarajan_et_al_2013,liu_et_al_2014a,liu_et_al_2014b}.  
Broadly speaking, there are two flavors of 21~cm power spectrum analysis: a ``delay
spectrum" approach, where the line-of-sight Fourier transform is done on individual
visibilities --- and is therefore not strictly orthogonal to the transverse directions because
of the frequency dependence of an individual visibility --- and an ``imaging" approach, where the line-of-sight Fourier transform spans multiple visibilities and is truly orthogonal
to the transverse directions on the sky.\footnote{
The terminology of an ``imaging" power spectrum is potentially misleading, but it has become somewhat standard
in the community.  The key feature is not that an image of the sky is made, 
but rather that visibilities are gridded
into the $uv$ plane and the frequency Fourier transform is taken in a direction 
truly orthogonal to $u$ and $v$.  The nomenclature of an ``imaging" power spectrum arises because
the gridded $uv$ data is only 2D spatial Fourier transform away from an image.}
A full discussion of the differences between these two approaches is outside the scope
of this work, but previous
analyses have shown that the wedge and the mapping from foreground sky position to 
$k_{\parallel}$ modes of the power spectrum remains valid for \emph{both} frameworks.
In the delay spectrum approach, the chromatic dependence of an individual baseline
is completely preserved, so that all foreground emission at a given location maps
to a given $k_{\parallel}$ mode.
An imaging approach, however, removes the mapping between delay and sky position
by projecting out the frequency sine wave for a known geometric delay.
In an imaging power spectrum, frequency structure is dominated by the intrinsic spectra of the sources,
so that a significant amount of foreground emission maps to low $k_{\parallel}$
modes, reflective of their inherent (smooth) frequency spectrum.
However, the chromatic response of the interferometer still affects the
observed emission, leading to a wedge feature analogous to that of the delay spectrum approach,
but with more of the emission concentrated at low $k_{\parallel}$
\citep{morales_et_al_2012,dillon_et_al_2015}.
Because of the brightness of foreground emission, this wedge still dominates any 21~cm signal
in the modes it occupies.

Explorations of these wide-field effects in actual data have been more limited.
\cite{thyagarajan_et_al_2015a,thyagarajan_et_al_2015b} studied both simulated and actual
MWA observations using the delay spectrum technique and found an excellent match between
the two, demonstrating a good understanding of both foreground emission and the primary
beam of the MWA.  They also found that the foreshortening of baseline lengths when projected
towards the horizon creates sensitivity to diffuse emission normally resolved out on
longer baselines.  Diffuse foregrounds are bright enough that they can be detected
despite the small (but non-zero) response of the MWA element towards the horizon.
This led to what they dubbed the ``pitch-fork" effect, a foreground signature in delay
space where bright emission from within the main field of view appeared at low delays
and emission from the horizon at high delays.

This work studies similar effects using an imaging power spectrum approach
and will confirm that the sky-position to $k_{\parallel}$ mapping still holds.
We will also focus on the ability to \emph{subtract} foreground emission away from
the main field of view to lower the contamination in high $k_{\parallel}$ modes.
In this section, however, we use the delay-spectrum formalism 
\citep{parsons_and_backer_2009,parsons_et_al_2012b} 
to provide a general framework for understanding these effects.
We stress that the delay spectrum  provides a straightforward, pedagogical
way to interpret power spectrum results, since the wide-field chromatic effects
appear at first-order.  As argued in \cite{morales_et_al_2012,trott_et_al_2012,liu_et_al_2014a,liu_et_al_2014b}, and as will confirmed with data below, 
these wide-field effects are generic to all interferometric
21~cm experiments.

The basic premise of the delay-spectrum technique presented in 
\cite{parsons_et_al_2012b} is that the square of the frequency Fourier transform of a single
baseline's visibility spectrum (i.e. the delay spectrum, $\tilde V_b(\tau)$) 
approximates a measurement of the cosmological power spectrum (to within a 
proportionality factor):
\begin{equation}
\label{eq:pspec_approx}
|\tilde V_b(\tau)|^2 \propto P(k_{\perp},k_{\parallel}),
\end{equation}
where
\begin{equation}
\label{eq:dtransform}
\tilde{V}_b(\tau) = \int \mathrm{d}\nu~V_b(\nu)~e^{2\pi i\nu\tau} 
\end{equation}
is the delay spectrum, $\tau$ is delay, $\nu$ is frequency, $V$ is a
visibility, and the subscript
$b$ indicates that the visibilities are from a single baseline.  

Intuitively, this relation is well-motivated.  To a good approximation,
a single baseline $b$ probes a single
transverse scale, and thus a single $k_{\perp}$ mode.  And, since cosmological
redshifting of the 21~cm line maps observed frequencies into line-of-sight 
distances, the Fourier transform of the frequency spectrum approximates a range
of $k_{\parallel}$ modes.  
Put more succinctly, for an interferometer, baseline length $b$ maps to cosmological $k_{\perp}$
and delay $\tau$ maps to $k_{\parallel}$.

The power of this simple formalism is that we can now map the effects of the
primary beam, which enter into a visibility measurement in a well-known way,
to cosmological Fourier space and the power spectrum 
$P(k_{\perp},k_{\parallel})$. We begin with the form of a visibility
in the flat-sky approximation \citep{thompson_et_al_2001}\footnote{Although 
use of the flat-sky
approximation to derive a wide-field interferometric effect may seem 
ill-motivated, it greatly simplifies the math in this pedagogical treatment.
See \cite{parsons_et_al_2012a,parsons_et_al_2012b} and \cite{thyagarajan_et_al_2015a,thyagarajan_et_al_2015b} for a discussion
of the subtleties introduced by the curved sky into the delay
formalism.}:
\begin{equation}
V_b(\nu) = \int~\mathrm{d}l~\mathrm{d}m~A(l,m,\nu)~I(l,m,\nu)~e^{-2\pi i(ul+vm)},
\end{equation}
where $A$ is the primary beam, $I$ is the sky brightness distribution,
$l$ and $m$ are direction cosines on the sky, $\nu$ is frequency,
and $u$ and $v$ are the
projected baseline lengths on the ground plane measured in wavelengths.
We can rewrite this expression in terms of the geometric delay $\tau_g$
\citep{parsons_and_backer_2009}:
\begin{equation}
V_b(\nu) = \int~\mathrm{d}l~\mathrm{d}m~A(l,m,\nu)~I(l,m,\nu)~e^{-2\pi i\nu\tau_g},
\end{equation}
where
\begin{equation}
\label{eq:taudef}
\tau_g = \frac{\mathbf{b}\cdot\hat{s}}{c} = \frac{1}{c}(b_xl+b_ym),
\end{equation}
$\mathbf{b} \equiv (b_x,b_y)$ is the baseline vector measured in meters
(i.e. $\mathbf{u} \equiv (u,v) = \nu\mathbf{b}/c)$, and $\hat{s} \equiv (l,m)$.
Doing the delay-transform given by Equation \ref{eq:dtransform} gives us
a delay-spectrum:
\begin{equation}
\tilde{V}_b(\tau) = \int~\mathrm{d}l~\mathrm{d}m~\mathrm{d}\nu~A(l,m,\nu)~I(l,m,\nu)~e^{-2\pi i \nu(\tau_g - \tau)}.
\end{equation}
If we make the pedagogical assumption that both $A$ and $I$ are independent
of frequency, we can straightforwardly do the delay
transform integral\footnote{\cite{parsons_et_al_2012b} showed that both the frequency-dependence of $A$ and $I$ create
a convolving kernel, broadening the footprint of each delay mode.  The 
ramifications of this effect are discussed below, but only
complicate the pedagogical nature of the current analysis.}:
\begin{equation}
\label{eq:dspec}
\tilde{V}_b(\tau) = \int~\mathrm{d}l~\mathrm{d}m~A(l,m)~I(l,m)~\delta(\tau_g - \tau).
\end{equation}
Since Equation \ref{eq:taudef} relates the geometric delay $\tau_g$ to a
specific set of sky direction cosines $(l,m)$, the delta function selects
a subset of sky positions which contribute to each $\tau$ mode in the delay
spectrum, albeit with a baseline-dependent
non-trivial mapping between sky position and $\tau$.
It is always true, however, that sources which appear at high delays are 
those which are far
from the pointing center of the instrument (hence the name ``horizon limit"
given to the maximum delay a source can appear at in 
\citealt{parsons_et_al_2012b}).  
Following Equation \ref{eq:pspec_approx}, we can say:
\begin{equation}
\label{eq:pk_dspec}
P(k_{\perp},k_{\parallel}) \propto \left[\int\mathrm{d}l~\mathrm{d}m~A(l,m)~I(l,m)~\delta(\tau_g-\tau)\right]^2,
\end{equation}
where the length of baseline $b$ sets $k_{\perp}$,
and $\tau \propto k_{\parallel}$.
This analysis therefore implies that sources at large
delays (i.e. sources near the edges of the field of view,
by Equation \ref{eq:taudef}) contaminate the highest $k_{\parallel}$ 
modes of the wedge 
$(k_{\perp},k_{\parallel})$ space.
Although not always stated as directly,
this result was also found in
\cite{morales_et_al_2012,vedantham_et_al_2012,thyagarajan_et_al_2013} and \cite{liu_et_al_2014a}
using entirely independent formalisms.

\begin{figure}[ht!]
\centering
\includegraphics[width=3in,clip=True,trim=.5cm 0cm 0cm 0cm]{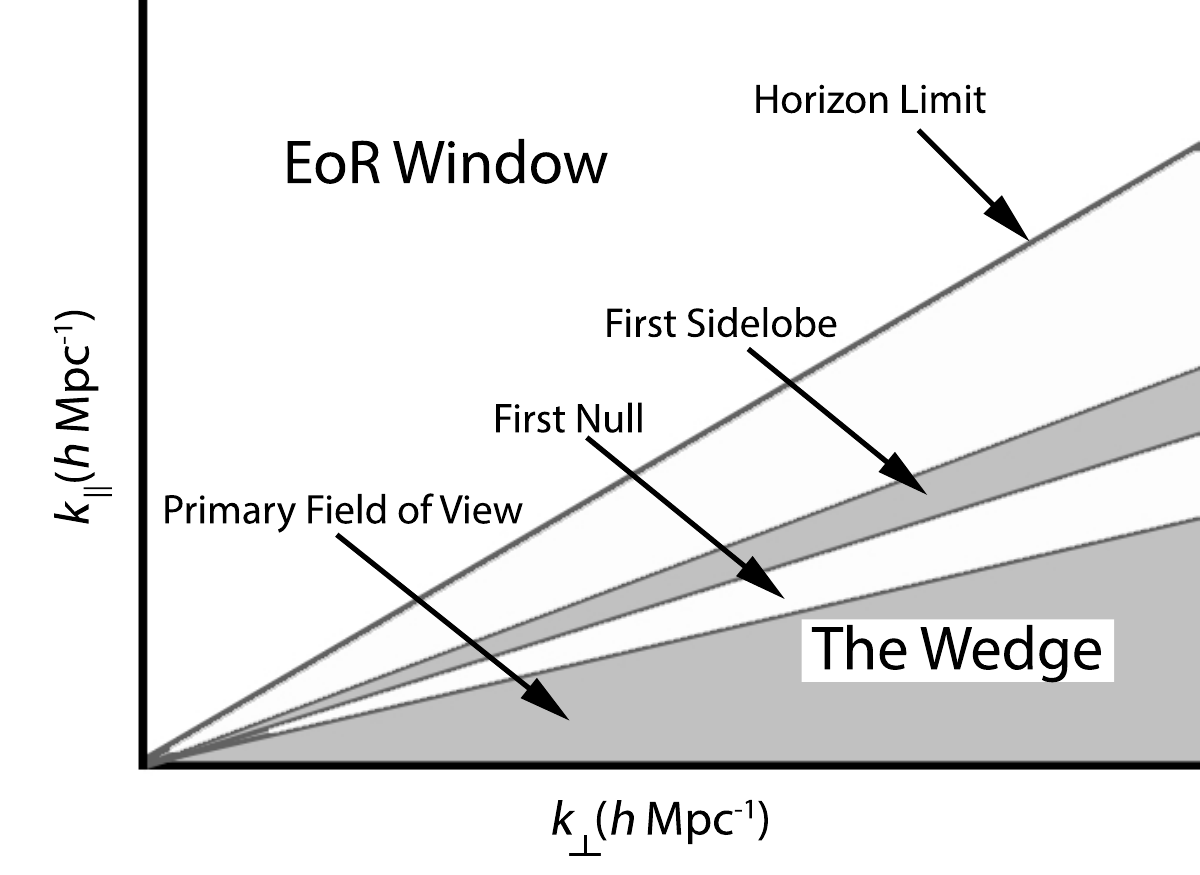}
\caption{A schematic diagram of the effects discussed here.
The primary beam attenuates foregrounds in the $k_{\parallel}$
direction.}
\label{fig:wedge-diagram}
\end{figure}

Equation \ref{eq:pk_dspec} also shows the main result we wished to derive 
in this section:
the (smooth spectrum) sky emission $I(l,m)$ which appears in each delay mode is
multiplicatively attenuated by the primary beam of the instrument.
Therefore, the foreground emission which contaminates those
$k_{\parallel}$ modes measured by a single baseline
will itself be attenuated by a (distorted) slice through the 
square of the primary beam
of the instrument.  
This result is schematically illustrated in Figure \ref{fig:wedge-diagram}.
Note that because delay space is a one-dimensional projection of
the sky coordinates (c.f. Equation \ref{eq:taudef}), the attenuating beam
in a $k_{\parallel}$ spectrum will vary depending on the orientation of the baseline.
On an east/west baseline, for example, the delay axis probes the relative east/west position of the source
and is insensitive to north/south translations in source positions.  Such a baseline will therefore
clearly show the effects of the eastern and western primary beam sidelobes in its $k_{\parallel}$
spectrum.  Similar logic applies to a north/south baseline and the northern and southern primary
beam sidelobes.  Delays on a 
northeast/southwest baseline, however, probe northeast/southwest sky position, and thus the east-west
translation of a source through the eastern and western sidelobes does not cause as rapid a change
in $k_{\parallel}$.  The net effect is that when all baselines of the same magnitude are averaged
into a $k_{\perp}$ bin, these different $k_{\parallel}$ sidelobe patterns add up and smear out
the location of the sidelobes.

An important but subtle point is that the above derivation for mapping
sky-coordinates into $k$ space was strictly for flat spectrum emission.
As shown in \cite{parsons_et_al_2012b}, any spectral structure
--- whether intrinsic to the source or the instrumental response --- 
introduces a convolving kernel that broadens the footprint of each
$k_{\parallel}$ mode in cosmological Fourier space.  While this kernel
is narrow for smooth-spectrum foregrounds, spectral structure in
the 21~cm signal spreads the 21~cm power across a wide range of $k_{\parallel}$
modes.  This is equivalent to saying that the 21~cm signal intrinsically has
power on these cosmological scales.  
The situation for foreground emission is different, however.
Although power spectrum plots are labeled
with axes of $(k_{\perp},k_{\parallel})$ with units of $h\rm{Mpc}^{-1}$,
these cosmological scalings apply \emph{only} to the 21~cm signal.  
The analysis
above shows how foregrounds map into this space, and how the primary beam
affects this mapping.
The primary beam of the instrument
does still act as a window function and can affect high $k_{\parallel}$ modes of the cosmological signal;
however, the cosmological signal has been shown to be relatively featureless on the scale of this
kernel (c.f. \citealt{parsons_et_al_2012b}), rendering this effect very small.
Regardless, the 21~cm signal is an all-sky signal with real
intrinsic $k_{\parallel}$ structure.  There is therefore always 21~cm signal
at the peak beam response, so there will always be power
at all $k_{\parallel}$ modes truly intrinsic to the cosmological signal.
This point will be discussed further in \S\ref{sec:discussion},
where we consider the possibility of detecting 21~cm emission at
$k_{\parallel}$ modes where the foregrounds fall in the nulls of the primary
beam.

The very wide and relatively smooth primary beam of the PAPER instrument
makes the predicted foreground attenuation difficult to see in the analysis of 
\cite{pober_et_al_2013b}.  However, for instruments like the MWA and LOFAR,
which use tiles of dipoles to increase the system gain and narrow the
size of the primary beam, there should be two clear effects visible in the
power spectra.  First, there should be significant attenuation of the wedge
foreground emission before the horizon limit, since the instrument field
of view is significantly smaller than $2\pi$ steradians,
as is seen in \cite{dillon_et_al_2014}.  Secondly,
at higher $k_{\parallel}$ values than those corresponding to
the main beam of the instrument, foreground emission should appear
coming from the sidelobes of the primary beam.  These two effects can be seen
in the delay-space simulations of different antenna elements presented in
\cite{thyagarajan_et_al_2015a}.
For an imaging power spectrum technique which averages baselines together, 
the second effect will be less clear 
for an instrument like the
MWA, in which all the dipoles and tiles are oriented in the same direction.
In this case,
the sidelobes are always oriented North/South and East/West;
as explained above, however, the beam footprint in $k_{\parallel}$,
will differ from baseline-to-baseline depending on that baseline's orientation
relative to sidelobe pattern.
This will have the effect of smearing out the sidelobe across a wider range
of $k_{\parallel}$ modes than would be seen in an instrument with circularly
symmetric sidelobes, but as we will show, the
feature is still quite visible in the power spectrum.

The structure of the remainder of this paper is as follows.
First, in \S\ref{sec:mwa}, we describe the MWA instrument and observations in more
detail.  With this context provided, we 
provide the results of two principal analyses.
In \S\ref{sec:sims}, we present simulated MWA power spectra made from a sky consisting of a
single point source.  By moving the position of this source from simulation to simulation,
we can see the primary beam effects described above in a controlled fashion.
In \S\ref{sec:data},
we use observations from the MWA to analyze these
primary beam features and present the power spectra of this data
in \S\ref{sec:pspec}.  In particular, we focus
on the effect of subtracting sources from 
sidelobes outside the primary field
of view.

\section{Observations with the Murchison Widefield Array}
\label{sec:mwa}

\begin{figure}[ht!]
\centering
\includegraphics[width=3.25in]{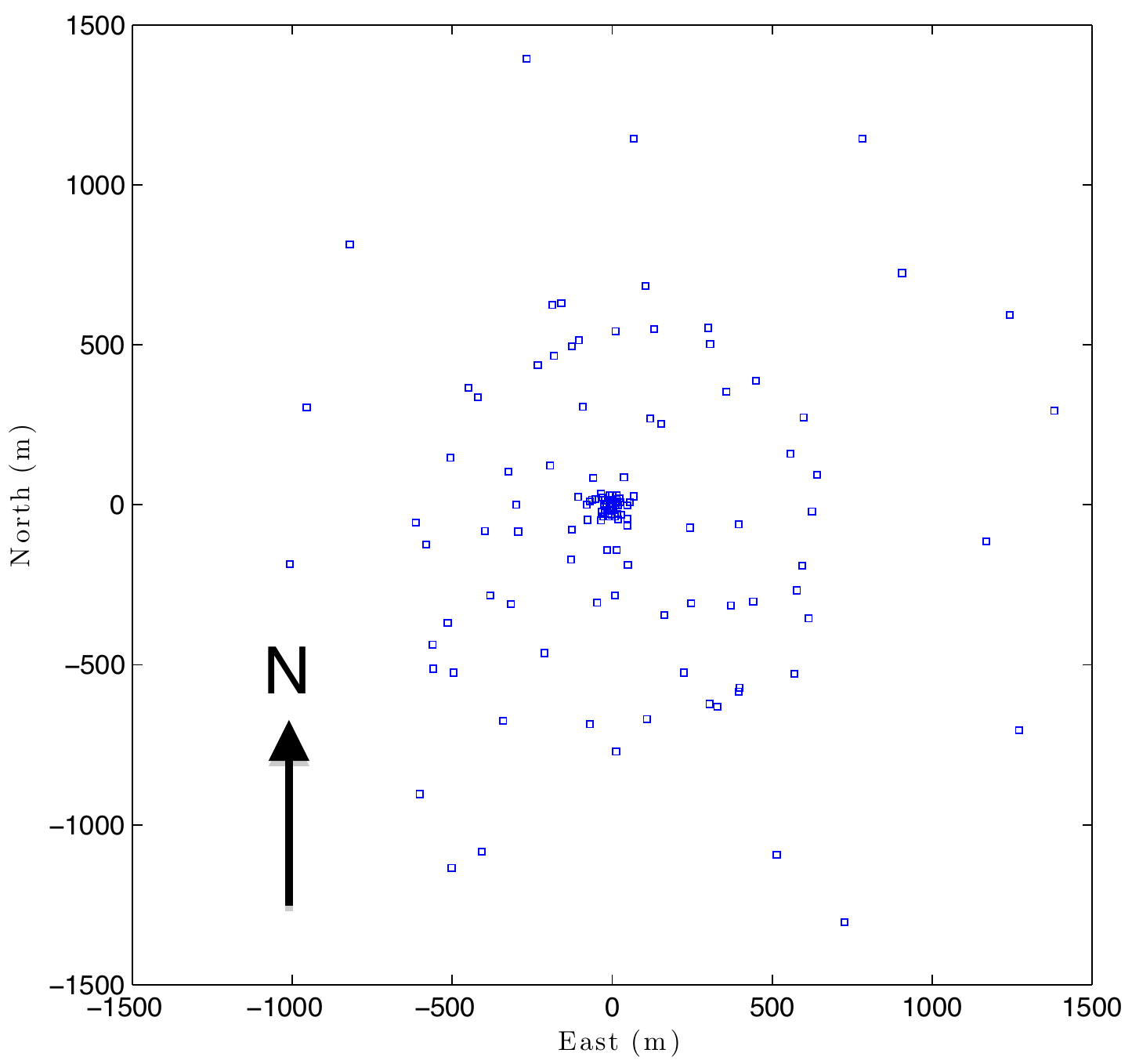}
\caption{MWA-128 array configuration; each square represents one tile
of 16 dipoles.}
\label{fig:arrayconfig}
\end{figure}

\begin{figure}[ht!]
\centering
\includegraphics[width=3.25in]{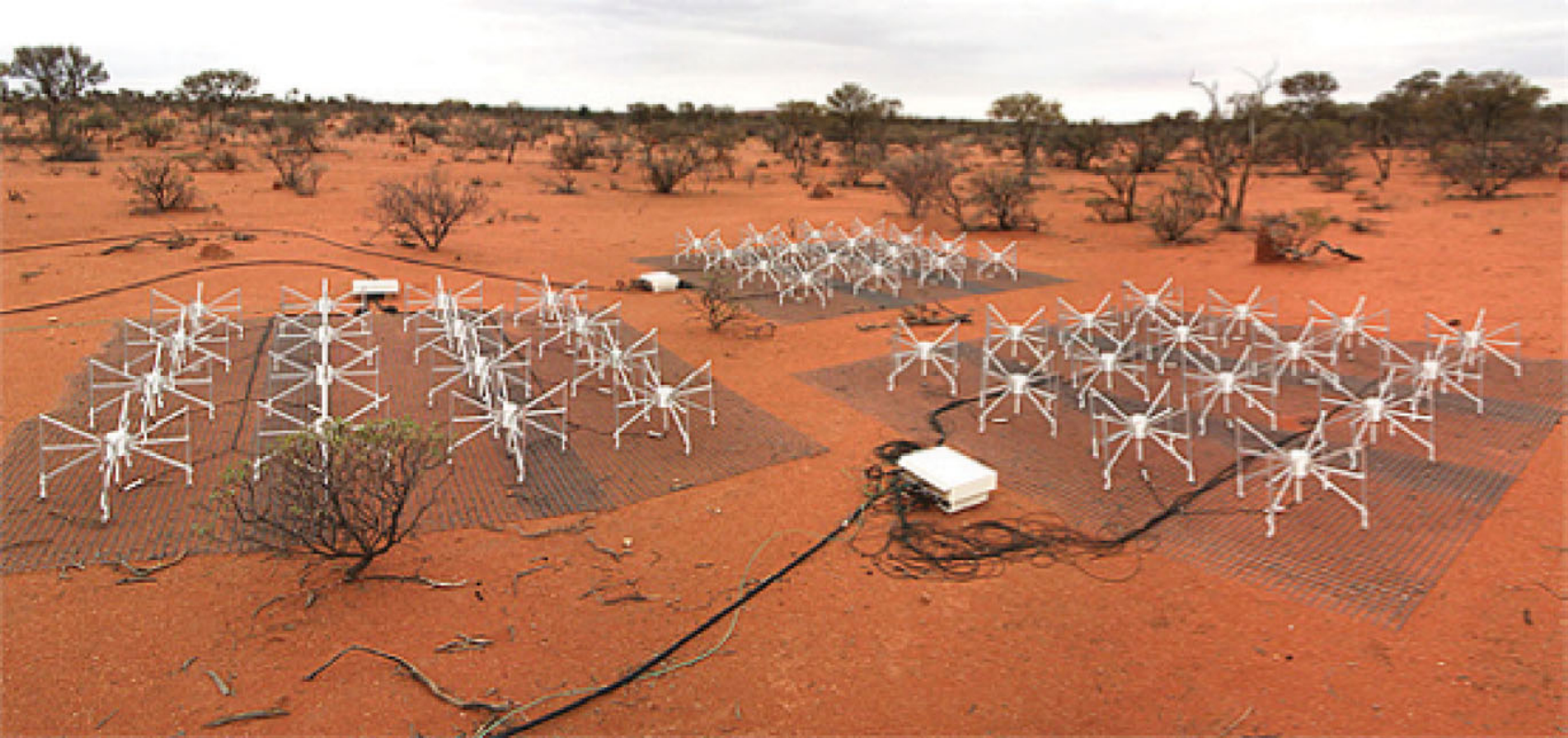}
\caption{Three MWA tiles, each consisting of 16 dual-polarization
dipole elements in a $4\times4$ grid.}
\label{fig:tile}
\end{figure}

The Murchison Widefield Array in Australia consists of 
128 tiles antenna elements, and each tile is composed of 16 dual-polarization dipole antennas;
the array configuration is shown in Figure \ref{fig:arrayconfig}.
The tile element has the effect of significantly narrowing the MWA's field
of view over that from a single dipole, but also introduces significant
regular sidelobes in the primary beam.  Figure \ref{fig:tile} shows three 
MWA tiles; every tile is aligned North/South, 
so the sidelobes from each tile
appear with nearly the same orientation.

The data used in this work was taken with the MWA on 23 Aug. 2013 (Julian Date
2456528) over the course of approximately 
three hours from 16:47:27 to 19:58:24 UTC.  
The observations were taken over a frequency band centered on 182.415~MHz,
with a total bandwidth of 30.72~MHz divided into 24 1.28~MHz coarse channels,
which are each further divided into 768~40~kHz fine channels.

\begin{figure}[ht!]
\centering
\includegraphics[width=3.5in]{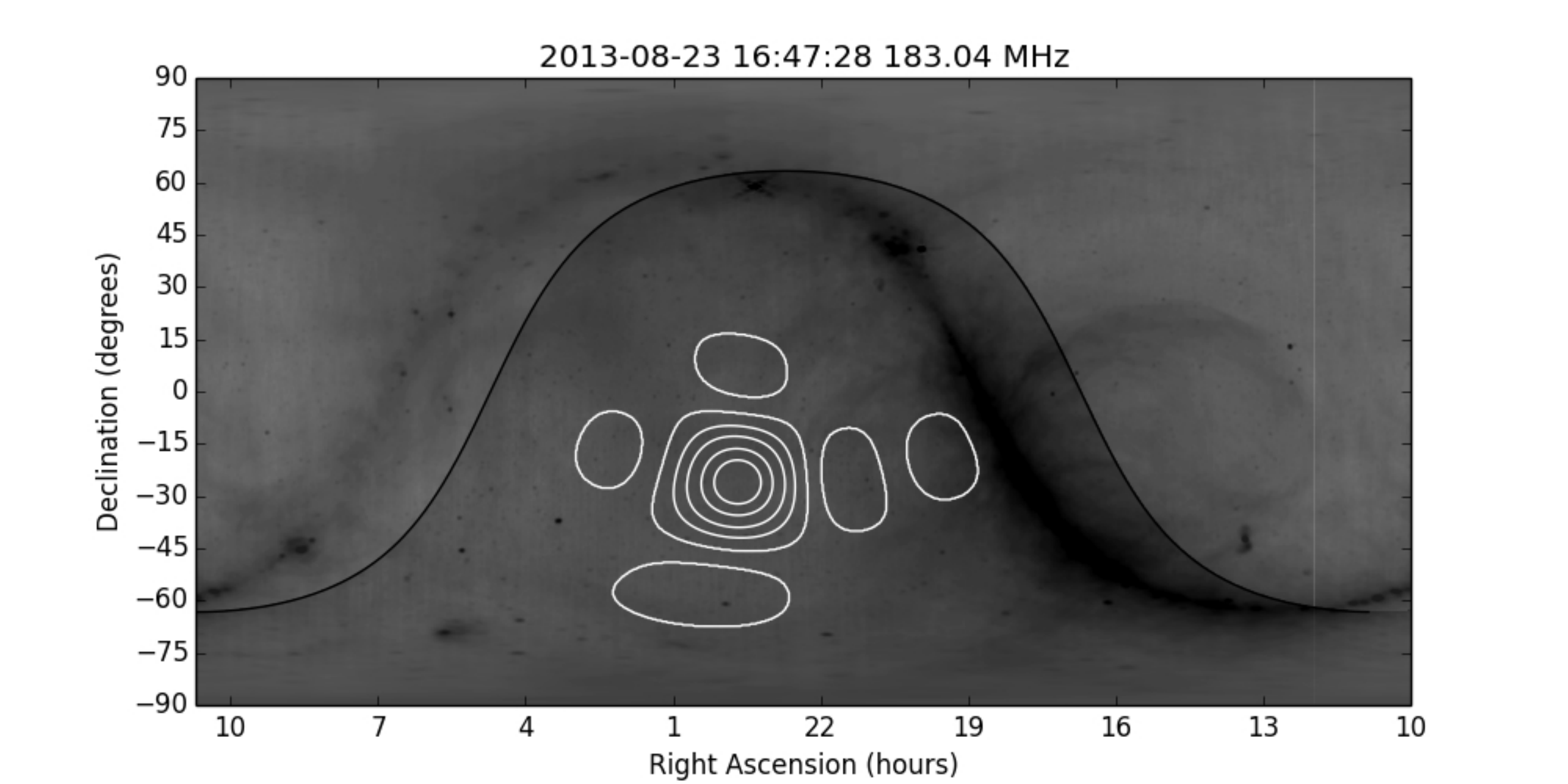}
\includegraphics[width=3.5in]{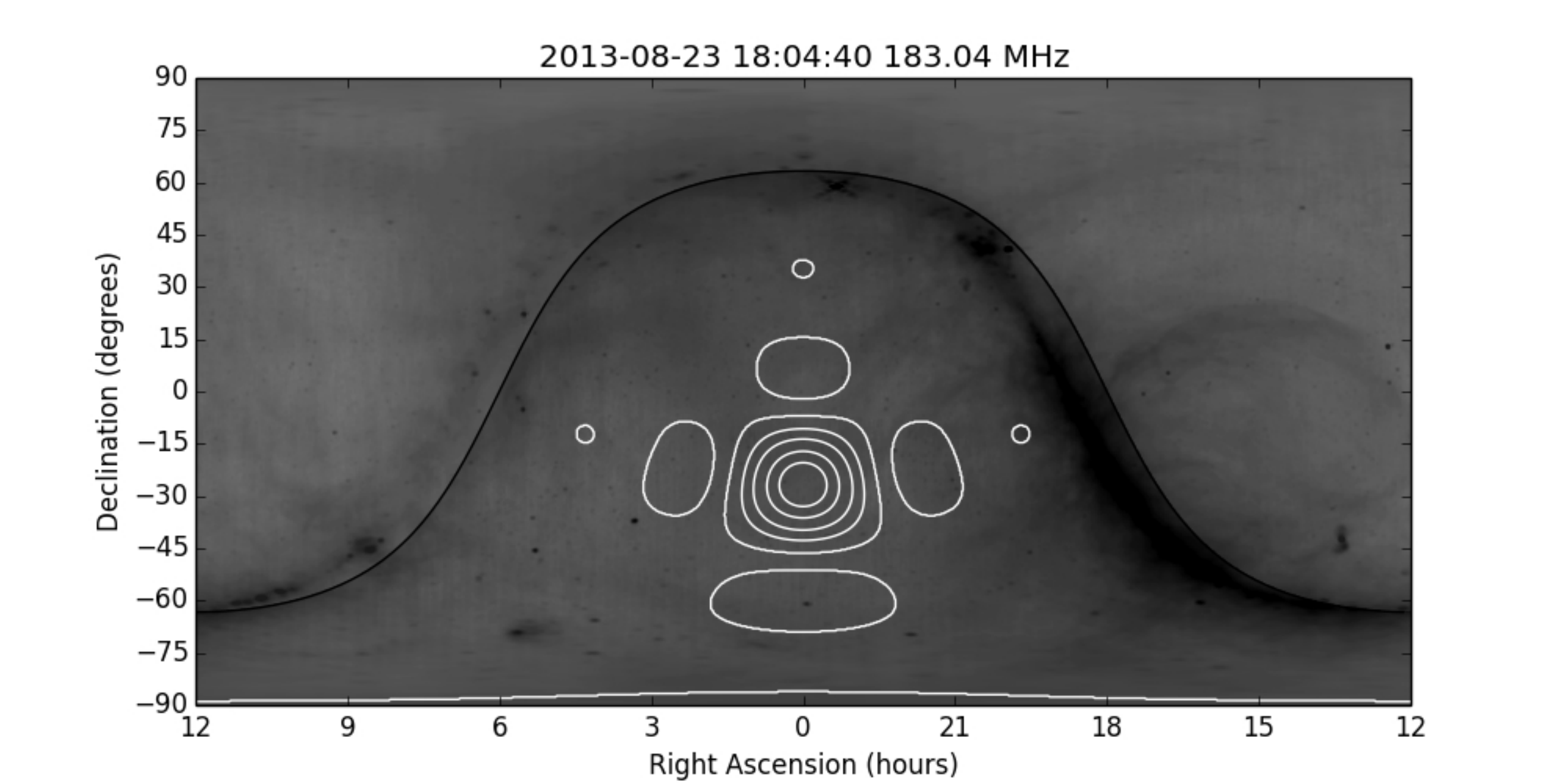}
\includegraphics[width=3.5in]{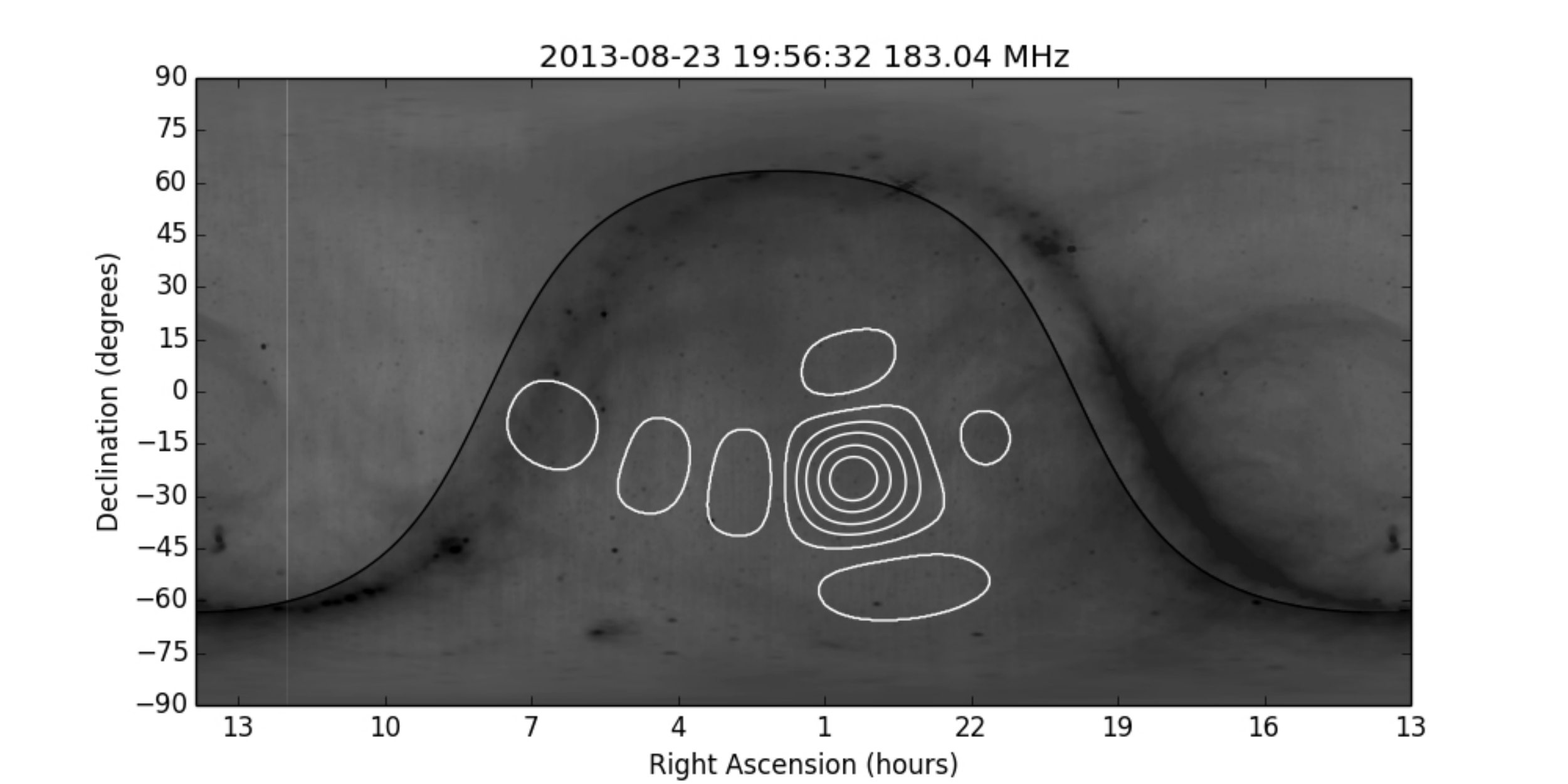}
\caption{Primary beam 
responses of the MWA tiles at several pointings.
White contours show the beam response over-plotted on the
\cite{haslam_et_al_1982} all-sky map; contour levels
are 0.01, 0.1, 0.25, 0.5, and 0.75 of peak beam response.
Although the sidelobes move over the course of the observation,
the main field of view remains relatively constant.
\emph{Top:} The first (earliest) pointing in the 3 hour data analyzed here.
\emph{Center:} The zenith-phased pointing near the center of the 3 hours.
\emph{Bottom:} The last (latest) pointing of the data set.
}
\label{fig:beams}
\end{figure}

The data used in this analysis span a total of 6 30 minute-long pointings, 
where an analog beamformer steers the main lobe of the primary beam to
nearly
the same sky coordinates for each pointing.  The sky is then allowed to drift
overhead for 30 minutes before re-pointing.  The data within
each pointing are saved as individual ``snapshot" observations, each
lasting 112 seconds, with individual integrations of 
0.5~s.
Figure \ref{fig:beams} shows the tile primary beam at three different
beamformer pointings: the beginning of the observation, a zenith-phased
pointing, and the end of the observation.
Since each pointing changes the
overall primary beam response of the instrument, the sidelobe patterns 
in the final integrated power spectrum will be smeared.
As will be shown below, however, the effects of the sidelobes are still
quite visible despite the changing primary beam shape.

\section{Pedagogical Simulations}
\label{sec:sims}

Before presenting the full analysis of this data set, we will first investigate the effects
of the location of celestial emission on the cosmological power spectrum and the wedge
in particular.  In this section, we will simulate visibilities for a single point source
and calculate the dependence of the power spectrum on the source's location.
Visibilities are simulated using the
Fast Holographic Deconvolution (FHD) software package.\footnote{Source code publicly available at https://github.com/miguelfmorales/FHD.}  Visibility simulation is one of several functions in FHD; 
as described below,
FHD also performs calibration and source subtraction on our actual data.
As a simulator, FHD constructs a $uv$ space model of the sky and integrates small
regions of the $uv$ plane using the holographic beam kernel \citep{morales_and_matejek_2009}
to create model visibilities.

For this analysis, we simulate visibilities for all the baselines in the MWA
in 768 fine frequency channels
spanning the observed 30.72~MHz frequency band.  We only simulate one 112~second 
snapshot when the primary
beam is pointed at zenith (i.e. the snapshot shown in the middle panel of Figure \ref{fig:beams}).
In addition to reducing the computational demand of the simulations, using only one snapshot
allows us to see the sidelobes patterns most clearly, since integrating over a longer amount of time
means including data when the array had a different pointing and primary beam.

\begin{figure}[ht!]
\centering
\includegraphics[width=3in]{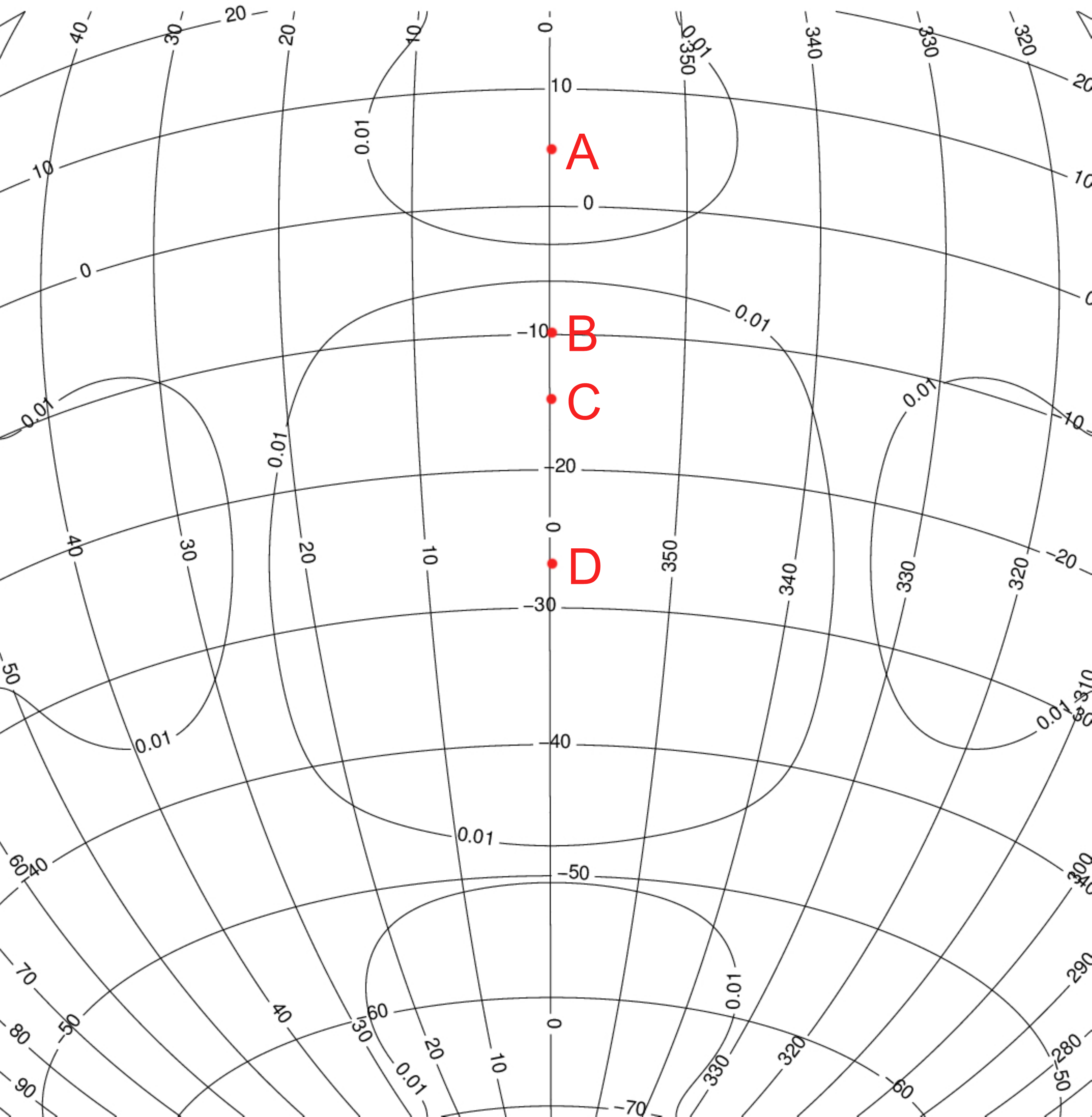}
\caption{Positions of the four sources simulated.  Source locations are in red; black
contours show the $1\%$ primary beam levels.  Note that there are four independent simulations,
each consisting of one point source only.  Letters correspond to the power spectra in Figure
\ref{fig:sim_pspecs}.}
\label{fig:sim_sources}
\end{figure}

We conduct four simulations, each consisting of one radio point source at a different location
on the sky; the locations simulated are shown in Figure \ref{fig:sim_sources}.
For each simulation, the inherent flux density of the source is increased relative to source D
(located at zenith) by the inverse of the primary beam response at its location.  In other words,
each source simulated has the same \emph{apparent} flux density.  This choice places all
the final power spectra on the same scale, allowing for easier comparison.

\begin{figure}
\centering
\includegraphics[height=8.25in]{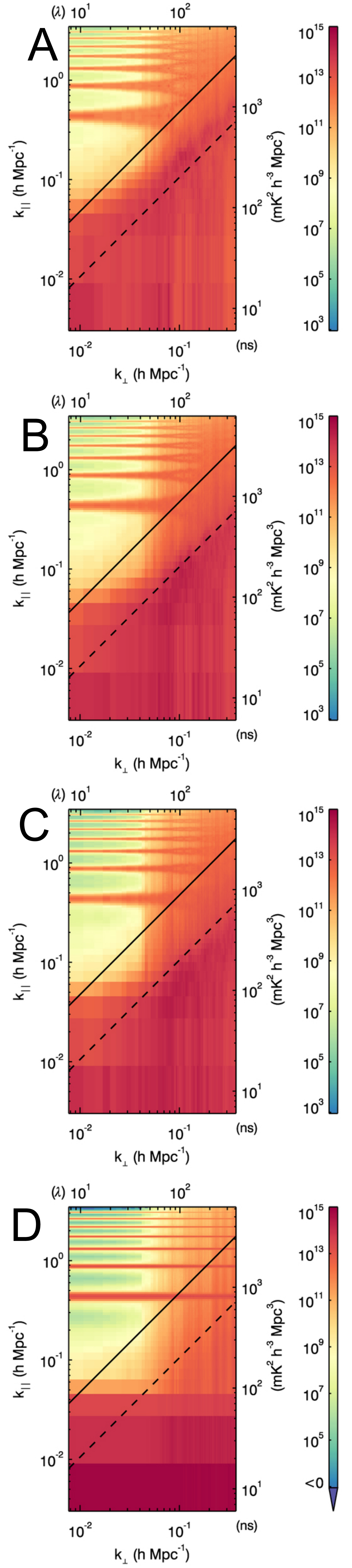}
\caption{
$(k_{\perp},k_{\parallel})$ power spectra of the simulated point sources.
Letters correspond to source positions in Figure \ref{fig:sim_sources}.
The solid black line shows the horizon limit; the dashed black line indicates the 
main field of view.
The wedge feature is absent for source D, located exactly at zenith, and power moves
higher in $k_{\parallel}$ as the source moves further from the center of the field of view.
Note that the schematic Figure \ref{fig:wedge-diagram} is plotted with linear axes, 
whereas this Figure uses logarithmic axes, 
which cause the horizon and field of view lines to be parallel.}
\label{fig:sim_pspecs}
\end{figure}

In Figure \ref{fig:sim_pspecs}, we show the 2D $(k_{\perp},k_{\parallel}$) power spectra
for each of the four simulations described above.  We show only
 the power spectrum from one of the two linear polarized dipoles
of the MWA; the power spectrum for the other polarization are quite similar.  
Letters correspond to the source labels
in Figure \ref{fig:sim_sources}.
To make the power spectrum, the simulated visibilities are imaged by FHD and then analyzed by
the $\epsilon$ppsilon pipeline described in Hazelton et al., (in prep.).\footnote{Source code publicly available at https://github.com/miguelfmorales/eppsilon.}
For more information on the data products transferred between FHD and $\epsilon$ppsilon,
see Jacobs et al., (in prep.).

The effect of source position on the power spectrum is clear and agrees with the intuition
developed in \S\ref{sec:formalism}.  Source D is located directly at zenith,
with the subsequent sources offset to higher declination (with right ascension held fixed).
In Figure \ref{fig:sim_pspecs}, source D exhibits no wedge feature.  (The power at high
$k_{\perp}$ values is due to poor $uv$ coverage on these scales and is described in more detail below.)
Sources C and B show a clear wedge feature arise as the source is moved away from zenith,
and the power spectrum of source A --- where the source is located in the sidelobe
of the primary beam --- shows a concentration of power outside the main field of view (indicated
by the dashed black line) but inside the horizon limit (solid black line).  This feature
is in exact accord with our predictions.
Simulations using sources offset in right ascension (instead of declination) show the same effect,
as do sources with offsets in both right ascension and declination: power moves to higher $k_{\parallel}$
as the source moves further from field center.

\section{Data Analysis}
\label{sec:data}

In this section we present the full analysis of the three hours of MWA data described in
\S\ref{sec:mwa}.  The data is processed through the same imaging and power spectrum analyses
(done by FHD and $\epsilon$ppsilon, respectively) applied to the simulations.  However,
there are initial pre-processing, calibration, and foreground subtraction steps applied to the data,
which we describe here.

\subsection{Pre-processing}

Pre-processing of the data uses the custom-built Cotter pipeline, which performs
time averaging of the integrations to 2~s and frequency averaging of the
narrow band channels to 80~kHz
\citep{offringa_et_al_2015}.
Cotter also uses the \textsc{aoflagger} code to flag and remove RFI
\citep{offringa_et_al_2010,offringa_et_al_2012}.  Cotter
also performs a bandpass correction, removing the spectral shape
within each coarse channel as well as correcting for variations in digital
gain between the coarse channels.
Finally, the data are converted from an MWA-specific data format
to \texttt{uvfits} files.

\subsection{Calibration and Imaging}

After the pre-processing, data are further calibrated and imaged using the
FHD software package.
FHD was designed for interferometers with wide fields of view
and direction dependent gains like the MWA and uses the holographic
beam pattern to grid visibilities to the $uv$ plane.  FHD also keeps track of the
gridding statistics in the $uv$ plane to allow for 
full propagation of errors through the image and into the power
spectrum.  

In this analysis, we do not use FHD to perform a deconvolution
and construct a source model from the data itself as was
described in \cite{sullivan_et_al_2012}; rather, we input
a catalog of point sources 
and use FHD to calculate model visibilities.  In all calculations, FHD uses 
a simulated primary beam model including the effects of mutual coupling between
dipoles in a tile \citep{sutinjo_et_al_2015}.

FHD also applies a calibration to the data, using the source model
provided to solve for frequency-dependent,
per-tile, per polarization complex gain parameters.  
Using an iterative approach, we reduce the number of free parameters by averaging the calibration
solutions into a bandpass model that is updated on a per-pointing (i.e. 30 minute) basis.  Depending on the position of a
tile in the array, one of five different length cables is used to return the signal for central processing;
we find it necessary to calculate a different bandpass model for each type of cable in the system.  
We also fit and remove a per-antenna polynomial (quadratic in amplitude, linear
in phase) that varies on a per-snapshot (112 second) timescale, as well as
a fit for a known ripple caused by a reflection within a 150~m cable.  
This particular cable is not present in all tiles, so the ripple is only removed from those
which contain this cable; reflections from cables of other lengths on other tiles appear to have
much smaller amplitude, although work is in progress to remove these effects as well.

For the present analysis, we image each snapshot at each frequency channel and make
3D image cubes in HEALPix \citep{gorski_et_al_2005}.
Each snapshot cube is then summed in image space to make a final integrated cube for power
spectrum analysis.

\subsection{Foreground Subtraction}

\begin{figure*}[ht!]
\centering
\includegraphics[width=7in]{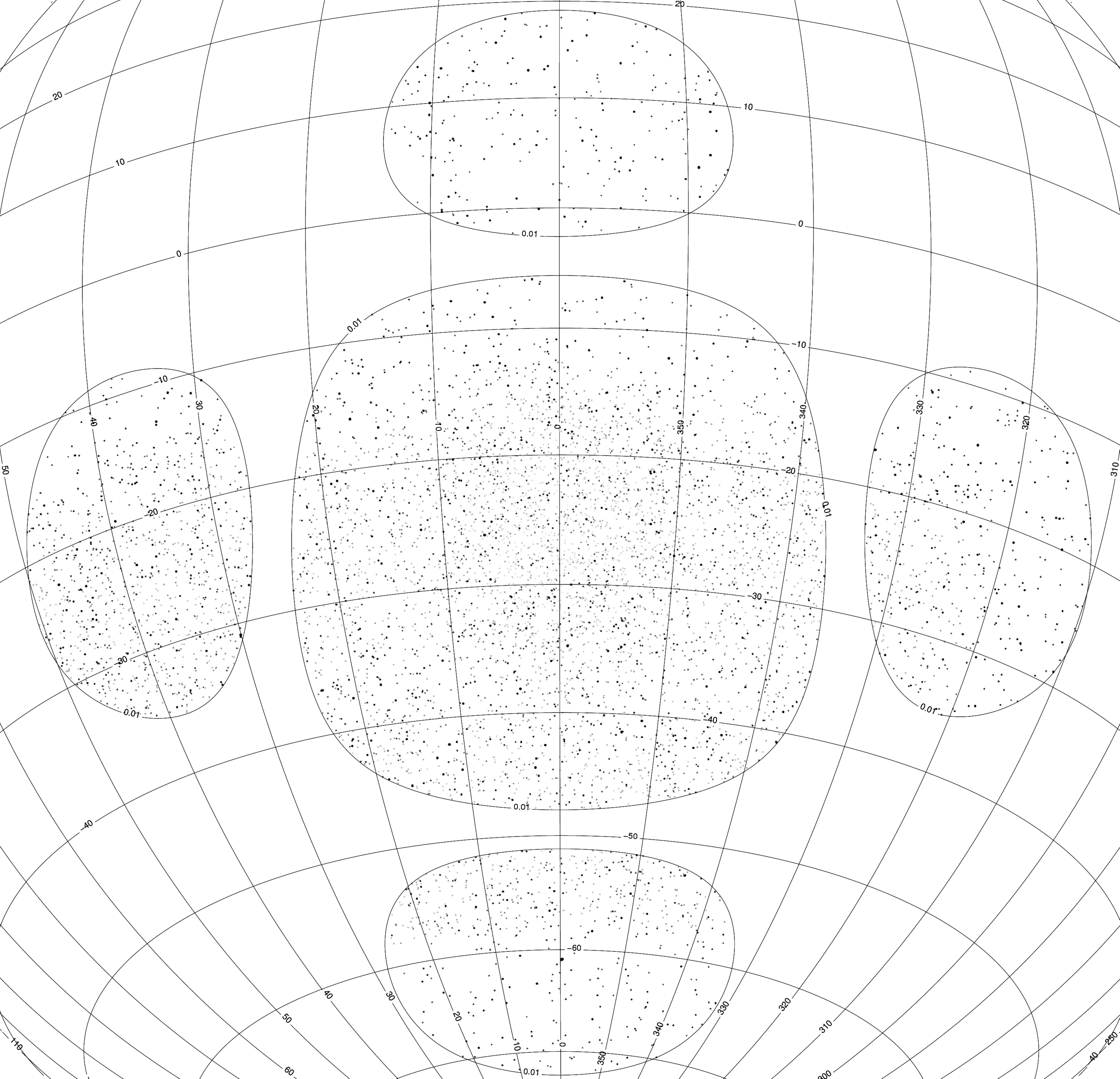}
\caption{The sources used for calibration and subtraction.  This image
shows the source positions during the zenith-phased pointing.  Any sources
where the beam response is greater than 1\% of the peak value during the
zenith pointing are included in our model.  The sidelobes are clearly
distinguishable from the main beam. The declination range of the MWACS survey
is the $-10^{\circ}$ to $-55^{\circ}$, which accounts for the drop in source density outside
this interval.}
\label{fig:map}
\end{figure*}

It is through FHD that model visibilities are also subtracted
from the data.  
We use two sets of model visibilities generated from
a custom-made point source catalog. 
In the main field of view, the catalog 
contains sources generated from FHD deconvolution outputs and an advanced machine-learning source identifier
designed to reject spurious sources (Carroll, et al., in prep.).
Outside the main field of view, our catalog combines sources from
MWA Commissioning Survey (MWACS; \citealt{hurley-walker_et_al_2014}), the Culgoora catalog \citep{slee_1995},
and the Molongolo Reference Catalog \citep{large_et_al_1981}
In one model we only include $\sim 4600$ sources 
that fall within the primary lobe of the 
MWA beam; in the other, we include all sources out through the first sidelobe
($\sim 8500$ sources).
An image of all the sources included during the zenith-phased snapshot is
shown in Figure \ref{fig:map}.
There are two effects that serve to limit the number of sources
included in our model.  First, we use a primary beam
threshold cut: any sources that fall where the beam response
is less than $1\%$ of the peak response are not included in the model.
Second, because it is a composite of several surveys, 
the completeness of our catalog is not uniform over the sky.
In particular, MWACS does not cover the full declination range of the
observations here; the effect is that fewer sources are removed from the 
lower declinations of the southern sidelobe, and very few are in the northern sidelobe.
This has the effect of introducing a small time-dependence in the number
of sources included in our model, since the declination coverage of the primary
beam does change with pointing (c.f. Figure \ref{fig:beams}).
MWACS also avoids the Galactic plane, which reduces the number of sources
in the model at the early and late pointings to $\sim 7000$.

It is also important to note that our sky
model assumes a fixed spectral index of -0.8 for each source.  Although
the actual sources on the sky will have some spectral structure, the
fact that we include minimal frequency-dependence in the model serves to
strengthen the arguments below: subtracting a nearly achromatic foreground 
model removes power from chromatic (i.e. high $k_{\parallel}$) modes of
the power spectrum.  This is a clear demonstration of the
inherent chromaticity of the interferometer response pattern.

\section{Power Spectra}
\label{sec:pspec}

\begin{figure*}[ht!]
\centering
\includegraphics[width=7in]{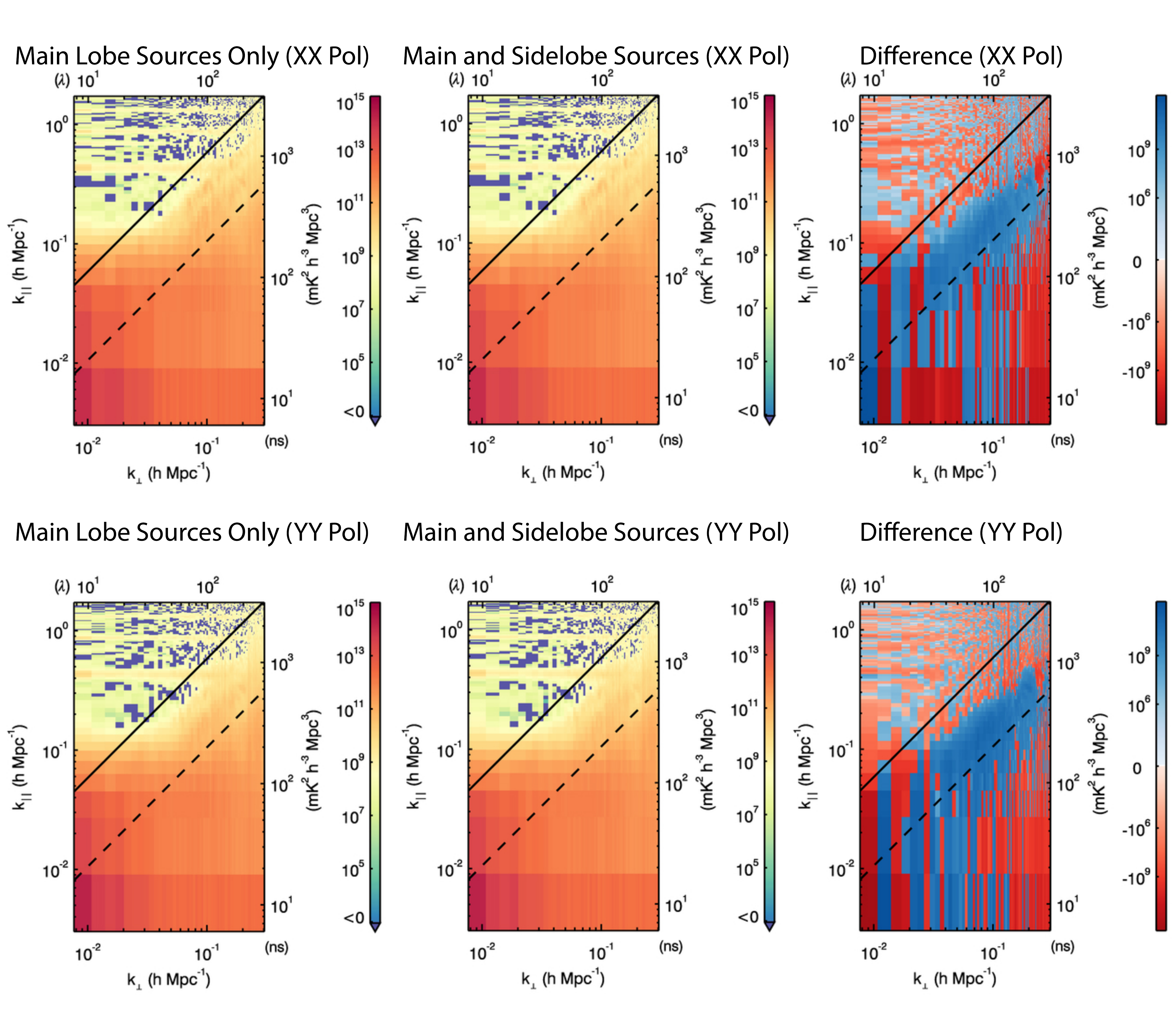}
\caption{$(k_{\perp},k_{\parallel})$ power spectra of the data.  XX linear
polarization is on the top row, YY on the bottom.  The solid black line
shows the horizon limit; the dashed black line indicates the 
main field of view.
\emph{Left}: Power spectra where only sources in the main lobe
of the beam are used to calibrate and then subtracted from the data.
\emph{Center}: Power spectra where sources in both the main lobe
and the sidelobes are used to calibrate
and then subtracted.
\emph{Right}: The difference between the left and center plots.  (Note
the data are differenced in 3D $(k_x,k_y,k_z)$ space and then averaged
in $k_{\perp}$ annuli.)  Although the left and center panels
appear indistinguishable, subtracting them reveals a significant difference
outside the first null of the primary beam.  The consistently blue
region shows that removing sources in the sidelobes has removed power
at high $k_{\parallel}$ outside the main field of view.
}
\label{fig:pspec}
\end{figure*}

We now present the power spectra of this data generated by the 
$\epsilon$ppsilon code.  With observational data, $\epsilon$ppsilon
empirically calculates the noise level in the visibilities and 
fully propagates errors in the visibilities
through to the 3D power spectrum.  
The important results here are the
cylindrically averaged 2D power spectra, shown in Figure \ref{fig:pspec}.
In this figure, the left hand panel shows the power spectrum with only
sources in the primary lobe removed, while the center panel shows
the power spectrum where sources are also subtracted from the sidelobes.
In order to enhance the subtle difference between the two panels,
we subtract the power spectrum including sidelobe source removal
from the power spectra which removes only main lobe sources (i.e.
we subtract the center panel from the left-hand panel).  Note
that we perform the subtraction in full 3D $(k_x,k_y,k_z)$ space
before binning into 2D $(k_{\perp},k_{\parallel})$ space.
We plot the result of this subtraction in the right hand panel of 
Figure \ref{fig:pspec}.  Most of the difference
randomly fluctuates between positive (blue) and negative (red) values, 
showing no systematic change of the power spectrum in these regions.  
However, the consistently blue
region shows that subtracting sources from the sidelobes
removes a non-trivial amount of power (as much as 10\% compared
to the power spectra with no sidelobe source subtraction, although typical
values are $\sim1\%$) from the region where
the sidelobe is expected: outside the main lobe (dashed black line)
but within the horizon (solid line).  Since the size of the main
lobe is frequency and pointing dependent, 
the dashed black line is only an approximate
marker; the power that is removed from $k_{\parallel}$ modes
below this line is consistent with being sidelobe power from
a range of frequencies and pointing centers.

Although not the primary goal of this paper, there are a few additional features
in the power spectra that warrant explanation.
\begin{itemize}
\item The horizontal lines running across the EoR window are the effect
of the coarse channelization used by the MWA.  Between each 1.28~MHz coarse
channel are two 80~kHz channels which are flagged due to low signal response
and potential aliasing concerns.
This flagging in frequency has the effect of introducing covariance into
the line-of-sight $k_{\parallel}$ modes, which are effectively
produced by a Fourier transform of the frequency axis.  This additional
covariance has the effect of coupling power from the wedge into higher
$k_{\parallel}$ modes.  Because the flagging is at regular intervals,
this additional power also appears at regular intervals in $k_{\parallel}$
(the appearance of non-regular spacing comes from the logarithmic scale on
the y-axis).  Work is underway on algorithms which can reduce this covariance
using priors on the fact that the power comes from $k_{\parallel}$ modes
within the wedge.

\item The vertical lines, which are especially prevalent at high $k_{\perp}$
modes come from the $uv$ coverage of the MWA.  The MWA has exceptionally
dense coverage at low $k_{\perp}$ due to its large number of short baselines.
However, at higher $k_{\perp}~(\gtrsim~10^{-1}~h{\rm Mpc}^{-1})$ 
there are gaps in the coverage, which results
in particularly noisy measurements of certain modes.  Therefore, while
these modes appear to have very high power, they also have associated
very large error bars.  A plot of the errors calculated by $\epsilon$ppsilon
for the XX polarization is shown in Figure \ref{fig:xx_errs}.
\begin{figure}[ht!]
\centering
\includegraphics[width=3in]{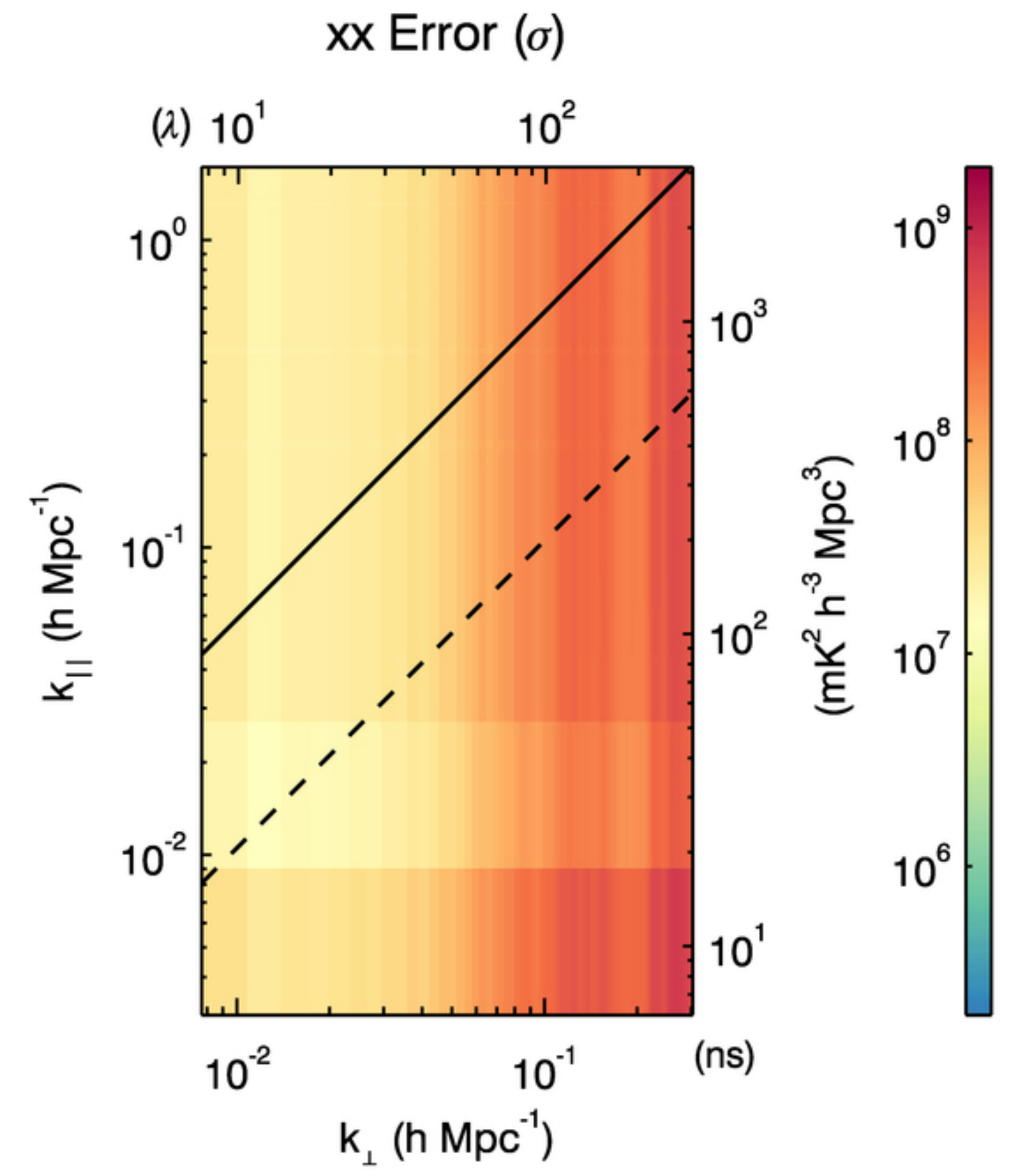}
\caption{The errors on the XX power spectrum shown in the upper left panel of Figure
\ref{fig:pspec} calculated by $\epsilon$ppsilon.  $uv$ coverage is worse at high $k_{\perp}$,
leading to higher errors.  These errors downweight the vertical streaks seen
at high $k_{\perp}$ in Figure \ref{fig:pspec} when estimating a 1D power spectrum.
}
\label{fig:xx_errs}
\end{figure}

\item There are blue/purple regions outside the wedge which are negative.  This
is because $\epsilon$ppsilon cross-multiplies the even time
samples in the data set with the odd time samples
(with the samples interleaved on a time scale of two seconds); this has the effect of
removing the positive-definite noise bias that would result from
squaring the entire data set.  Alternating positive and negative values
correspond to noise dominated regions. 

\item Most obviously, a large amount of foreground power remains in the power
spectra.  This is not surprising, as our analysis only subtracted a
few thousand point sources, ignoring diffuse emission both from the
Galaxy and unresolved point sources.  Subtracting models of this emission
will clearly be necessary for any possibility of recovering 21~cm signal
from inside the wedge.  The effects concerning sidelobes
presented in this work, however, are still
quite important: the additional fraction of emission removed when
including sidelobe sources is more than enough to swamp the EoR signal
which might have a peak power spectrum brightness 
on order of $10^6~{\rm mK}^2h^{-3}{\rm Mpc}^3$.

\end{itemize}

\section{Discussion}
\label{sec:discussion}

Through the advances in our understanding of EoR foregrounds (i.e.
the ``wedge" and ``EoR window" paradigm), we now have a model for the
detailed impact of sources far from pointing center
on the recovery of the 21~cm power 
spectrum.  This work demonstrates that sources outside the main
field of view are a significant contaminant of the modes of interest
in the 21~cm power spectrum, even for an ``imaging" power spectrum analysis.
It is therefore worthwhile to heuristically
consider the detailed pattern sources far from pointing center 
leave in cylindrically
averaged ($k_{\perp},k_{\parallel}$) space.  
While not all of the conclusions below directly follow from 
the empirical power spectra analyzed
here, the formalism presented in \S\ref{sec:formalism},
the delay spectrum analyses in \cite{thyagarajan_et_al_2015a,thyagarajan_et_al_2015b}
combined with the results of our sidelobe source subtraction from MWA 
observations suggest several interesting lines of reasoning.
To guide this discussion, 
we divide the sky
into three rough categories --- the primary field of view, the sidelobes,
and the nulls --- and discuss the effects of sources appearing in each
regime.

\begin{itemize}
\item{\textbf{Primary Field of View}}: These are the sources that are
traditionally considered when treating foreground removal from 21~cm
experiments.  Although the primary field of view may not be at zenith
for phased array telescopes (e.g. MWA and LOFAR), phase rotation
still places these sources at low delays, and therefore, low $k_{\parallel}$
in power spectrum measurements.  The dashed line in Figure \ref{fig:pspec}
roughly
indicates the edge of the main lobe of the MWA, and the brightness
of emission can be seen to clearly fall as one moves to higher
$k_{\parallel}$ modes.
Since this emission is located where the beam response is
at a maximum, it appears as the brightest contaminant in the
21~cm power spectrum. 
Because they are detected at high signal-to-noise, point sources in
the primary field of view are often used to simultaneously calibrate the
instrument response while they are subtracted from a visibility model
(e.g. \citealt{yatawatta_et_al_2013}).  Diffuse emission and unresolved point
sources generally dominate the total foreground power, requiring additional
models or parametric methods for removal (e.g. \citealt{chapman_et_al_2012}).

\item{\textbf{Sidelobes}}:
As seen in the present work, emission in the sidelobes appears at
higher $k_{\parallel}$ values than emission from inside the primary field
of view.  Therefore, a model of the sidelobes and the emission that falls
within them must be subtracted from the data in order to recover these
$k_{\parallel}$ modes closer to the EoR window.  
The primary beam attenuation has the effect of reducing the
fractional accuracy required in modeling these sources, since only their residual
apparent flux density contaminates the power spectrum    
Since emission in sidelobes contaminates higher
$k_{\parallel}$ modes than emission in the main beam, though,
removing wide-field emission may be more effective
at reducing leakage into the EoR window than removing emission from the
primary field of view.

\item{\textbf{Nulls}}: Lastly, one might assume that the nulls in the primary
beam might serve as good ``EoR windows", just like the regions 
of $k$ space outside the horizon.  
(Recall the discussion in \S\ref{sec:formalism}: while nulls
on the sky clearly attenuate all emission from that sky position;
however, when we refer
to beam nulls in $k_{\parallel}$, these nulls are confined to the foreground
emission.  EoR signal from other positions on the sky still produces
unattenuated power at these $k_{\parallel}$ modes.)
And while
the present analysis does indeed suggest that areas of $k$ space
corresponding to sky positions of exceedingly low beam response
will be free from foreground contamination, some caveats must be issued.
First, the nulls between sidelobes are likely not as deep as analytic
models suggest, due to effects like mutual coupling between the dipoles
in a station and group delay errors \citep{neben_et_al_2015}.
Second, as derived in \cite{parsons_et_al_2012b},
there is a non-negligible $k$ space point spread function convolving each source of emission.
Therefore, while there may be narrow nulls between sidelobes, emission
within the sidelobes can contaminate these nulls due to this spillover
effect.  Finally, it is worth remembering that the mapping from zenith
angle to delay (which, recall, maps to $k_{\parallel}$) is non-linear.
A delay bin near the horizon corresponds to much lower elevation
that a delay bin near zenith.  This means that while emission in bins 
far from the main lobe of the beam are strongly attenuated by the beam 
response, the total aggregate sum of emission in that $k_{\parallel}$ bin
can still be large, since it corresponds to a large area of sky
\citep{thyagarajan_et_al_2015a}.
It may still be that instruments like SKA and HERA with 
narrower fields of view and less sensitivity to foreground
emission away from pointing center
(\citealt{thyagarajan_et_al_2015a} estimate SKA and HERA will suppress of emission near 
the horizon 40~dB more than MWA),
the only safe place for foreground ``avoidance" is
beyond the horizon.

\end{itemize}

These arguments have important ramifications for experiments looking
to subtract foreground emission and recover $k$ modes from inside the wedge.
In particular, they suggest 
that experiments looking to ``enlarge" the EoR window
and remove foreground emission from modes near the horizon will benefit
most from subtracting emission outside the main lobe.
For this subtraction to be effective, accurate
wide-field primary beam calibration is necessary to properly
characterize the sidelobe patterns as a function of frequency.  Such
wide-field calibration may require new techniques, e.g., 
\cite{pober_et_al_2012,yatawatta_et_al_2013,neben_et_al_2015}.
Additionally, these arguments motivate the need for low-frequency, wide-field
sky surveys, especially in the Southern Hemisphere where the vast majority
of EoR frequency 21~cm experiments are being constructed.  Experiments
like HERA do not have a steerable beam, and thus will have difficulty
measuring the flux densities of sources in its sidelobes.  An accurate catalog
produced by another survey covering a larger area (e.g. \citealt{jacobs_et_al_2011,williams_et_al_2012,hurley-walker_et_al_2014,wayth_et_al_2015}) will be highly
valuable for subtracting sources outside the main field of view.  
Northern Hemisphere
experiments like LOFAR and GMRT may also be valuable for characterizing
the foregrounds at higher declinations.

Another important conclusion of this work is the implication that foreground
subtraction cannot simply target the removal of some total amount of flux 
density independent of the position
of that emission on the sky.  
Even if all emission from inside the primary field of view could be
perfectly removed,
sources in the sidelobes will continue to dominate higher 
$k_{\parallel}$ modes.  To recover all modes of the 21~cm power spectrum,
foreground models must extend into any primary beam sidelobes where
the level of beam attenuation does not reduce the foreground
power below that of the EoR signal.
While attention has been paid to the removal of bright off-axis point
sources \citep{offringa_et_al_2012}, the remaining diffuse emission and
confused source background will still have significant spectral structure
from the instrumental effects we have described.  Given the extremely
high foreground-to-signal ratio in 21~cm experiments, emission far from
the pointing center
cannot be neglected even if it is largely attenuated by the instrument
primary beam.

In practice, wide-field source subtraction at the level needed to recover
the EoR signal may require more than an accurate foreground model,
especially for experiments with very wide fields of view like PAPER.
Firstly, curved sky effects become important near the horizon;
an imaging based analysis that does not correctly handle the curved
sky (e.g. with $w$-projection; \citealt{cornwell_et_al_2008}) could
wash out the input source model and reduce the amount of power subtracted.
Secondly, ionospheric effects may become important for the level of
accuracy needed in subtraction 
\citep{mitchell_et_al_2009,bernardi_et_al_2009,intema_et_al_2009}.  The
ionosphere may introduce frequency-dependent, time-dependent, and
direction-dependent gains, all of which could lead to errors
in model-based source subtraction if not corrected for.  None
of these effects alleviate the need for wide-field foreground subtraction,
however; rather, they increase the difficulty of implementing a scheme
that potentially remove foregrounds to a level below the EoR
signal.

\section{Conclusions}
\label{sec:conclusions}

In this work, we have presented a heuristic description of the imprint
of the primary beam in power spectrum measurements from 21~cm interferometers.
In particular, we find that wide-field effects --- especially primary
beam sidelobes --- leave a highly chromatic imprint, so that even
smooth spectrum emission that falls within the sidelobes corrupts
high $k_{\parallel}$ modes of the power spectrum.  We further demonstrate
this effect both with pedagogical simulations using single point sources and
by removing a source model from
MWA observations.  When the model includes sources out to the first
primary beam sidelobes, it produces a significant (percent level) 
reduction in power
at high $k_{\parallel}$ values.

This result has significant implications for experiments looking
to measure the power spectrum of 21~cm emission from the EoR or any other
epoch.  In particular, it shows that foregrounds \emph{must} be considered
as a wide-field contaminant.  Only removing foregrounds from the primary
field of view will not reduce power in high $k_{\parallel}$ modes corresponding
to the primary beam sidelobes.  As a corollary, pipeline simulations which
only include foregrounds within the primary field of view are missing
a major contaminant of the EoR signal.  Experiments looking to 
use foreground subtraction to enlarge the EoR window must also pay
particular attention to emission in the sidelobes, since these sources
are those that corrupt modes closest to the EoR window.

\acknowledgments{The authors wish to thank Adrian Liu for helpful 
conversations and our referee for a number of helpful suggestions that improved the paper.
JCP is supported by an NSF Astronomy and Astrophysics 
Fellowship under award AST-1302774.
This scientific work makes use of the Murchison Radio-astronomy Observatory, operated by CSIRO. We acknowledge the Wajarri Yamatji people as the traditional owners of the Observatory site. Support for the MWA comes from the U.S. National Science Foundation (grants AST-0457585, PHY-0835713, CAREER-0847753, and AST-0908884), the Australian Research Council (LIEF grants LE0775621 and LE0882938), the U.S. Air Force Office of Scientific Research (grant FA9550-0510247), and the Centre for All-sky Astrophysics (an Australian Research Council Centre of Excellence funded by grant CE110001020). Support is also provided by the Smithsonian Astrophysical Observatory, the MIT School of Science, the Raman Research Institute, the Australian National University, and the Victoria University of Wellington (via grant MED-E1799 from the New Zealand Ministry of Economic Development and an IBM Shared University Research Grant). The Australian Federal government provides additional support via the Commonwealth Scientific and Industrial Research Organisation (CSIRO), National Collaborative Research Infrastructure Strategy, Education Investment Fund, and the Australia India Strategic Research Fund, and Astronomy Australia Limited, under contract to Curtin University. We acknowledge the iVEC Petabyte Data Store, the Initiative in Innovative Computing and the CUDA Center for Excellence sponsored by NVIDIA at Harvard University, and the International Centre for Radio Astronomy Research (ICRAR), a Joint Venture of Curtin University and The University of Western Australia, funded by the Western Australian State government.}

\bibliographystyle{yahapj}
\bibliography{ms}{}

\end{document}